\documentclass[useAMS,usenatbib,usegraphicx]{mn2e}

\usepackage{aas_macros,url,amssymb}

\title[Tibet's Ali: Asia's Atacama?]{Tibet's Ali: Asia's Atacama?}
\author[Quan-Zhi Ye et al.]{Quan-Zhi Ye$^{1}$\thanks{E-mail:
qye22@uwo.ca}, Meng Su$^{2,3}$, Hong Li$^{4}$, and Xinmin Zhang$^{4}$\\
$^{1}$Department of Physics and Astronomy, The University of Western Ontario,
London, Ontario, N6A 3K7 Canada\\
$^{2}$Kavli Institute for Astrophysics and Space Research, Massachusetts Institute of Technology, Cambridge, MA 02139, USA\\
$^{3}$Einstein Fellow\\
$^{4}$Institute of High Energy Physics, Chinese Academy of Science, Beijing 100049, China}
\begin{document}

\date{Accepted 1969 December 31. Received 1969 December 31; in original form 1969 December 31}

\pagerange{\pageref{firstpage}--\pageref{lastpage}} \pubyear{2016}

\maketitle

\label{firstpage}

\begin{abstract}
The Ngari (Ali) prefecture of Tibet, one of the highest areas in the world, has recently emerged as a promising site for future
astronomical observation. Here we use 31 years of reanalysis data from the Climate Forecast System Reanalysis (CFSR) to examine the astroclimatology of Ngari, using the recently-erected Ali Observatory at Shiquanhe (5~047~m above mean sea level) as the representative site. We find the percentage of photometric night, median atmospheric seeing and median precipitable water vapor (PWV) of the Shiquanhe site to be $57\%$, $0.8''$ and 2.5~mm, comparable some of the world's best astronomical observatories. Additional calculation supports the Shiquanhe region as one of the better sites for astronomical observations over the Tibetan Plateau. Based on the studies taken at comparable environment at Atacama, extraordinary observing condition may be possible at the few vehicle-accessible 6~000~m heights in the Shiquanhe region. Such possibility should be thoroughly investigated in future.
\end{abstract}


\section{Introduction}

At over 4~500 meters above mean sea level, Tibet's Ngari prefecture (also spelled as Ali) is one of the highest areas in the world. The high altitude, dry environment, minimal light pollution and radio interference has all made Ngari a promising site for next generation astronomical observatories. The National Astronomical Observatories of China (NAOC) has erected a permanent observatory at an elevation of 5~047~m near the town of Shiquanhe in 2010 \citep[c.f.][]{Yao2012}\footnote{See also \url{http://sitesurvey.bao.ac.cn/defaulte.html}, accessed 2015 Sep. 15. Shiquanhe, the largest settlement of Ngari and the seat of the Gar county, is also occasionally being confusingly referred to as Ali or Gar (the town). To avoid this confusion, we will use Ngari or Gar when referring to the prefecture or county and Shiquanhe when referring to the town.}. Several small telescopes are already in operation, with a few meter-sized telescopes in active development.

From the climatological aspect, the potential site for a future astronomical observatory is usually evaluated by its long-term cloudiness, atmospheric seeing, and the amount of precipitable water vapor. Long-term in-situ monitoring observation usually provides crucial information on site characterization, but it also takes considerable time and resources to complete. Alternatively, we take advantage of the service operation of numerical weather models to evaluate the astroclimatology of a potential site, which provides guidance and reference for future long-term in-situ observations. Here we present a detailed analysis of the astroclimatology of the Ngari region, using the Climate Forecast System Reanalysis (CFSR) data over a time span of 31 years, to assess the site for future astronomical observations.

\section{Methodology}

The Climate Forecast System (CFS) is a medium to long range numerical weather model developed by the National Centers for Environmental Prediction (NCEP). Besides routine forecast service that is being run 4 times a day, CFS has been used on reconstructing the history of the atmosphere using historic meteorological measurements, resulting the Climate Forecast System Reanalysis (CFSR), which covers a time span of 31 years from 1979 to 2010 \citep[c.f.][]{Saha2010v, Decker2012a}. CFSR is chosen over other reanalysis products because it has a better spatial resolution ($0.3^\circ$) and direct cloud amount output. Readers may refer to works such as \citet{Ebisuzaki2011}, \citet{Wang2011}, and \citet{Bao2013} for a review of the performance of the CFSR model.

The CFSR data is retrieved from the Research Data Archive managed by the Data Support Section of the Computational and Information Systems Laboratory at the National Center for Atmospheric Research \footnote{Accessible at \url{http://rda.ucar.edu}.}. To derive astroclimatology-related variables (namely cloudiness, atmospheric seeing and precipitable water vapor or PWV), we retrieve the data of the following variables:

\begin{enumerate}
 \item 6-hour average high level cloud cover at resolution of $0.3^\circ$ (equivalent to 30~km at the latitude of Ngari). High level cloud data (which corresponds to clouds with base height of $\gtrsim$5~000~m) is chosen over the ``entire atmospheric column'' cloud data, as the CFSR model is usually under-sampled over mountainous area of which fine geological features (like local plateau and peaks where astronomical observatories are often built upon) are unresolvable. The cloud amount is computed following the technique described in \citet{Xu1996f}.
 
 The CFSR cloud amount is then used to determine the observing condition. For the purpose of this work, we only consider data points that corresponds to local night time. A night is defined as ``photometric night'' if at least one data point of the night has a cloud amount of $<5\%$; the night is defined as usable if the cloud amount is $<50\%$ \citep[c.f.][]{Sarazin1990}.
 \item 6-hourly temperature at 2-m and 30~mb above surface, $0.995\sigma$ level, and 22 vertical levels from 900~hPa to 10~hPa, at resolution of $0.5^\circ$ except the 2-m data (which is at $0.3^\circ$). The temperature data is used to calculate the atmospheric seeing following the approach by \citet{Trinquet2006} and \citet{Ye2011}. The seeing is calculated only when the sky is deemed usable.
 \item 6-hourly precipitable water vapor (PWV) for the entire atmospheric column (at $0.3^\circ$). PWV refers to the depth of water in a column of the atmosphere and can be derived by integrating the water content along the atmospheric column in the respective grid box. However, due to the terrain under-sampling issue as outlined above, the actual PWV may be lower for high altitude sites.
\end{enumerate}

\section{Analysis}

The methodology is first verified with the reported astroclimatological values of established high altitude (over 4~000~m) astronomical observatories, including University of Tokyo Atacama Observatory at Cerro Chajnantor (5~640~m), the Cerro Tolonchar candidate site for European Extremely Large Telescope (E-ELT; 4~480~m), Indian Astronomical Observatory at Mount Saraswati near Hanle (4~467~m), and Mauna Kea Observatory (4~190~m). Quantities derived from CFSR data along with the long-term values are tabulated in Table~\ref{tbl-verify}. The following remarks are made from the comparison:

\begin{enumerate}
 \item The cloud amounts agree within $\sim 10\%$, while the CFSR values always equal or lower than the actual values. The cloud amounts are unaffected by the terrain under-sample issue as only high level clouds are considered.
 \item Almost all CFSR seeings are higher than the in-situ values; the magnitude of the deviation seems to be proportional to the difference between actual elevation and the CFSR grid elevation. For the case of Hanle which the actual elevation and the grid elevation has minimal difference, the CFSR seeing seems to agree with the actual observation.
 \item The CFSR PWV values are always higher than the in-situ values, most prominently for the case of Mauna Kea, possibly due to the steep terrain around the area \citep[see also][]{Suen2014}. The magnitude of the deviation are proportional to the difference between actual elevation and the CFSR grid elevation, similar to the behavior of the CFSR seeing.
 \item Overall, it seems that the CFSR values can be taken as a conservative estimate of the observing conditions, caution must be taken if the grid elevation differs too much from the actual elevation (such as the case of Mauna Kea), as the CFSR values are too conservative in such cases.
\end{enumerate}

\begin{table*}
\centering
\caption{Comparison between the long-term medians derived from the CFSR data and in-situ measurements for Cerro Chajnantor \citep{Giovanelli2001, Gio01a, Yoshii2010}, Cerro Tolonchar \citep{Schoeck2009}, Hanle \citep{Cowsik2002a} and Mauna Kea \citep{Morrison1973h, Bely1987c}. The medians derived from recent measurements in 2012--2014 \citep[as asterisked; see][]{Liu2015,Wang2015,Yao2015}, along with the calculated CFSR values for Shiquanhe Observatory, are given at the end of the table.}
\begin{tabular}{lccccccccccc}
\hline
 Site & Lon \& Lat & Elev. & Grid Box & \multicolumn{2}{c}{Percentage of} & \multicolumn{2}{c}{Percentage of} & \multicolumn{2}{c}{Seeing} & \multicolumn{2}{c}{PWV} \\
 & & & Elev. & \multicolumn{2}{c}{Phot. Nights} & \multicolumn{2}{c}{Usable Nights} & \multicolumn{2}{c}{} & \multicolumn{2}{c}{(mm)} \\
 & & & & Obs. & CFSR & Obs. & CFSR & Obs. & CFSR & Obs. & CFSR \\
\hline
 Co. Chajnantor & $67.76^\circ$~W, $23.02^\circ$~S & 5640~m & 4121~m & $63\%$ & $54\%$ & $82\%$ & $70\%$ & $0.7''$ & $1.3''$ & 0.5 & 1.8 \\
 Co. Tolonchar & $67.98^\circ$~W, $23.93^\circ$~S & 4480~m & 3201~m & - & $62\%$ & $82\%$ & $79\%$ & $0.6''$ & $1.1''$ & 1.7 & 3.0 \\
 Hanle & $78.96^\circ$~E, $32.78^\circ$~N & 4467~m & 4888~m & $52\%$ & $54\%$ & $71\%$ & $70\%$ & $0.8''$ & $0.9''$ & $<2$ & 2.5 \\
 Mauna Kea & $155.47^\circ$~W, $19.82^\circ$~N & 4190~m & 1488~m & $56\%$ & $59\%$ & $76\%$ & $67\%$ & $0.4''$ & $1.2''$ & 1 & 17 \\
\hline
 Ali Obs. (Shiquanhe) & $80.03^\circ$~E, $32.33^\circ$~N & 5047~m & 5004~m & $72\%$* & $57\%$ & $80\%$* & $75\%$ & $0.8''$* & $0.8''$ & - & 2.5 \\
\hline
\end{tabular}
\label{tbl-verify}
\end{table*}

\begin{figure}
\includegraphics[width=0.5\textwidth]{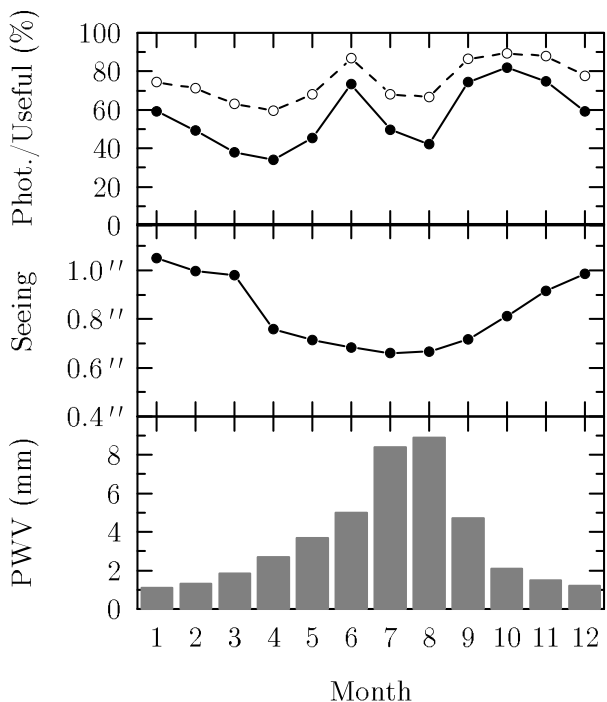}
\caption{From top to bottom: monthly variations of percentage of photometric (solid symbols and lines) and usable night (hollow symbols, dashed lines), atmospheric seeing and PWV of Shiquanhe Observatory, as derived from the CFSR data.}
\label{fig-detail}
\end{figure}

The computed values for Ali Observatory at Shiquanhe are appended to Table~\ref{tbl-verify}. Long-term measurements are currently unavailable, but the computed values grossly agree with the numbers derived from measurements taken in 2012--2014 \citep{Liu2015,Wang2015,Yao2015}. In general, the astroclimatology of Shiquanhe Observatory is in par with other sites listed in Table~\ref{tbl-verify}. On the other hand, the observing condition varies significantly throughout the season (Figure~\ref{fig-detail}): the best observing season in terms of percentage of photometric/usable nights is found in early highland spring (May to June) and late autumn into winter (September to January), separated by the rain season (July to August) and the wind season (late winter to early spring). Winter months have lower PWV (near 1.0~mm) while summer months tend to have better seeing conditions.

While the general observing condition at the Shiquanhe site appear to be decent, the possibility of extraordinary observing conditions at higher altitude must not be overlooked. The Shiquanhe region is not short of 6~000~m-class peaks and heights. A few of them are close to existing infrastructure, relatively gentle in slope, and are accessible by four-wheel-drive vehicle. These sites are unresolvable in the current CFSR model as they are almost a kilometer above the average grid elevation. Multi-altitude measurements performed by \citet{Giovanelli2001} at Atacama suggest that even modest elevation gain would be sufficient to achieve significantly better seeing ($1.1''$ vs. $0.7''$ from 5~000~m to 5~100~m) and PWV (1.2~mm to 0.7~mm from 5~000~m to 5~400~m, 0.5~mm at 5~750~m). The environment and climate of the Shiquanhe region are comparable to Atacama, hence it is possible that the observing condition at these 6~000~m sites are significantly better than the observing condition at 5~000~m.


\begin{figure*}
\includegraphics[width=\textwidth]{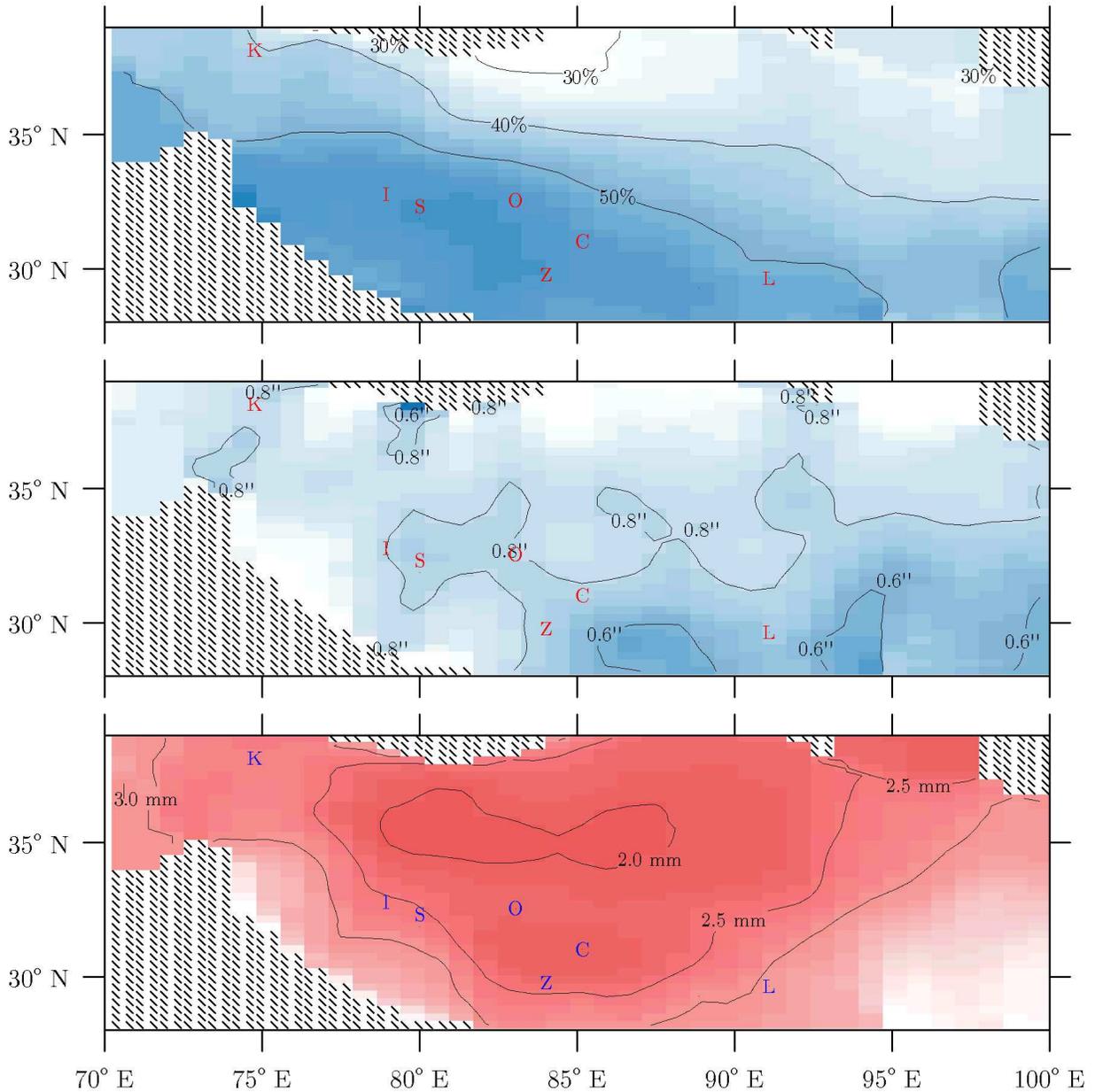}
\caption{CFSR's 31-year percentage of photometric night, atmospheric seeing and PWV across the Tibetan Plateau. Astronomical observatories, site testing monitoring posts and major inhabited settlements are marked: K -- Kalasu, I -- Indian Astronomical Observatory (Hanle), S -- Shiquanhe Observatory, O -- Oma, Z -- Zhongba, C -- Coq\^{e}n, L -- Lhasa.}
\label{fig-region}
\end{figure*}

The last question we look into is the overall astroclimatology of the Tibetan Plateau. To answer this question, the calculation is repeated for the grids between $70^\circ$ to $100^\circ$~E and $28^\circ$ to $39^\circ$~N with grid elevation over 4~500~m. As shown in Figure~\ref{fig-region}, the southwestern half of the Plateau, including Ngari, sees the highest percentage of clear weather, while the clearest weather falls slightly east and southeast of Shiquanhe, namely between Gar, G\^{e}rz\^{e} of Ngari and Zhongba of Xigaz\^{e}. The area with better-than-average seeing condition falls in central and eastern half of the Plateau, the latter of which is in an extremely rough terrain. The PWV is very low across the most part of the Plateau, with the minimum region falls over the northern highland of Ngari and Nagqu, as well as the Ngari-Xigaz\^{e} border between Zhongba and Coq\^{e}n, some 400~km east of Shiquanhe. Overall, the CFSR data supports the Shiquanhe region as one of the better sites for astronomical observations over the Tibetan Plateau. The Zhongba-Coq\^{e}n region may be another promising site, however we note that the terrain of this region is somewhat rougher than the Shiquanhe region, making it difficult to find suitable sites for infrastructure development.

\section{Summary}

We use the 31-year CFSR reanalysis data to examine the astroclimatology of Tibet's Ngari region as a potential site for future astronomical facilities. Three quantities are calculated from the CFSR data: percentage of photometric and usable nights, calculated from the cloud amount data provided by CFSR; atmospheric seeing, calculated by applying \citet{Trinquet2006} and \citet{Ye2011}'s model to CFSR's temperature data at multiple atmospheric layers; as well as PWV, which is directly provided by CFSR. We find the medians of percentage of photometric night, seeing and PWV to be 57\%, $0.8''$ and 2.5~mm for Shiquanhe Observatory in the core of Ngari, comparable to some of the world's best astronomical observatories. Based on these findings, we conclude that the Shiquanhe site has the potential for future optical, infrared and millimeter-wavelength observations, and could potentially provide a unique site in the northern hemisphere for cosmic microwave background experiments complimentary to existing sites at Antartica and Atacama. Additional calculations also support the Shiquanhe region as one of the better sites for astronomical observations over the Tibetan Plateau, while the Zhongba-Coq\^{e}n region, located some 400~km east of Shiquanhe, appears to be another promising site.

In addition, we note that the few vehicle-accessible 6~000~m-class heights in the Shiquanhe region may worth close attention. These sites are not resolvable in the CFSR model, but may be capable to provide extraordinary observing conditions, as hinted by multi-altitude experiments at comparable environment at Atacama. Such possibility should be thoroughly examined in future.

\section*{Acknowledgments}

We thank an anonymous referee for his/her comments. We also thank Yong-Qiang Yao, Li-Yong Liu, Yong-Heng Zhao (National Astronomical Observatories of China); Ji Yang, Sheng-Cai Shi (Purple Mountain Observatory); Xiaowei Liu (Peking University); and Scott Paine, John Kovac (Harvard-Smithsonian Center for Astrophysics) for discussion. Q.-Z. thanks Paul Wiegert for providing computational resource.


\label{lastpage}

\end{document}